\documentclass[aps,prc,amsmath,amssymb,floatfix,twocolumn,superscriptaddress]{revtex4-2}
\usepackage{graphicx}
\usepackage{amsfonts}
\usepackage{color}
\usepackage[colorlinks=true,linkcolor=blue,citecolor=blue]{hyperref}
\usepackage{dcolumn}
\usepackage{bm}
\usepackage{braket}

\usepackage[normalem]{ulem}  

\begin{document}

\title{
Five-dimensional collective Hamiltonian with improved inertial functions
}

\author{Kouhei Washiyama}
\email[E-mail: ]{washiyama@nucl.ph.tsukuba.ac.jp}
\affiliation{Center for Computational Sciences, University of Tsukuba, Tsukuba 305-8577, Japan}
\affiliation{Research Center for Superheavy Elements, Kyushu University, Fukuoka 819-0395, Japan}
\author{Nobuo Hinohara}
\affiliation{Center for Computational Sciences, University of Tsukuba, Tsukuba 305-8577, Japan}
\affiliation{Faculty of Pure and Applied Sciences, University of Tsukuba, Tsukuba 305-8571, Japan}
\author{Takashi Nakatsukasa}
\affiliation{Center for Computational Sciences, University of Tsukuba, Tsukuba 305-8577, Japan}
\affiliation{Faculty of Pure and Applied Sciences, University of Tsukuba, Tsukuba 305-8571, Japan}
\affiliation{RIKEN Nishina Center, Wako 351-0198, Japan}

\date{\today}

\begin{abstract}
\begin{description}
\item[Background] 
To describe shape fluctuations associated with large-amplitude collective motion in the quadrupole degrees of freedom, the five-dimensional collective Hamiltonian (5DCH) has been widely used.
The inertial functions in the 5DCH are microscopically calculated
with the energy density functional (EDF) theory employing the cranking formula.
However, since the cranking formula ignores dynamical residual effects,
it is known to fail
to reproduce the correct inertial functions, for instance, the total mass for the translational motion.

\item[Purpose]
We aim to resolve problems of the insufficient description of the inertial functions in the 5DCH.
We provide a practical method to include the dynamical residual effects in the inertial functions 
that depend on the quadrupole deformation parameters $\beta$ and $\gamma$.

\item[Methods]
We use the local quasiparticle random-phase approximation (LQRPA)
based on the constrained Hartree-Fock-Bogoliubov states in the $\beta$--$\gamma$ plane
with the Skyrme EDF.
The finite-amplitude method 
is used for efficient computations of the LQRPA.

\item[Results]
The inertial functions evaluated with the LQRPA 
significantly increase from the ones with the cranking formula due to the dynamical residual effects. 
This increase also shows a strong $\beta$--$\gamma$ dependence.
We show an application of the present method to a transitional nucleus $^{110}$Pd.
The low-lying positive-parity spectra are well reproduced with the LQRPA inertial functions.

\item[Conclusions]
We clarify the importance of the dynamical residual effects in the inertial functions of the 5DCH
for the description of the low-lying spectra.
The 5DCH with the improved inertial functions
provides a reliable and efficient description of low-lying spectra in nuclei
associated with the quadrupole shape fluctuation.

\end{description}
\end{abstract}

\maketitle

\textit{Introduction.}
%
A proper and feasible description of the shape dynamics in the ground and the excited states is
one of the important subjects in nuclear physics.
%
Observations of spectroscopic properties in nuclei 
suggest the existence of shape fluctuations and shape coexistence phenomena in low-lying states in nuclei,
particularly in the so-called transitional regions
from spherical to deformed shapes in the nuclear chart \cite{heyde11}.

%
The self-consistent nuclear energy density functional (EDF) theory has often been employed to describe ground-state properties of nuclei \cite{bender03,nakatsukasa16}.
To describe
shape fluctuations and shape coexistence phenomena associated with large-amplitude collective motion, 
it is necessary to use beyond-mean-field methods.
The generator-coordinate method (GCM)
with the quadrupole deformation parameters $\beta$ and $\gamma$ as generator coordinates
\cite{bender08,yao10,rodriguez10,yao14,kimura12,suzuki21}
has been developed and shown the importance of including the triaxial degree of freedom, $\gamma$.
Recently, 
the standard GCM was extended
to construct the basis states stochastically \cite{shinohara06,fukuoka13} and
variationally \cite{matsumoto23}.
%
Although the GCM is a fully quantum theory,
in practice, we need to combine the GCM with the projection method to recover the broken symmetries.
The GCM with the projection method requires a large amount of numerical computations.
In addition,
there 
remain many unsolved issues 
with realistic EDFs \cite{nakatsukasa16}.
For instance, the discontinuities and divergences are 
caused by the fractional powers of the density dependence
in EDFs \cite{anguiano01,dobaczewski07}.


As an alternative approach to the GCM,
the five-dimensional collective Hamiltonian (5DCH) method \cite{bohr, prochniak09} 
with the intrinsic quadrupole deformation parameters $(\beta, \gamma)$ and the three Euler angles
has been extensively used based on the EDF
\cite{libert99,prochniak04,prochniak09, niksic09,delaroche10}.
In most of the EDF-based 5DCH studies,
the inertial functions in the vibrational and rotational kinetic energies
are calculated according to the formula in the adiabatic perturbation \cite{ring-schuck},
which is identical to the well-known Inglis-Belyaev (IB) formula for the rotational moment of inertia \cite{inglis56,beliaev61}.
The vibrational masses are further approximated by the so-called perturbative cranking formula \cite{girod79}.
The cranking formula ignores variation of the self-consistent potential induced by the collective motion,
known as the dynamical residual effects,
thus, giving an insufficient description of the inertial functions \cite{dobaczewski81}.
In particular, the absence of the time-odd terms of the dynamical mean field leads to
the violation of the Galilean symmetry and is known to produce the wrong translational mass
\cite{wen22}.
Despite such drawbacks,
the cranking formula has been widely used \cite{libert99,prochniak04,prochniak09, niksic09,delaroche10},
because the full inclusion of the dynamical residual effects
in the inertial functions 
requires a huge computational cost.
Some recent 5DCH studies \cite{delaroche10,delaroshe24} evaluate the rotational moments of inertia within the cranked Hartree-Fock-Bogoliubov (HFB) framework
that are equivalent to the Thouless-Valatin inertia
\cite{thouless62} 
to include the dynamical residual effects in the rotational kinetic energy.
%
In many of the former studies,
a phenomenological enhancement factor of 1.2--1.4 is adopted for the inertial functions
evaluated with the cranking formula.


To properly include the dynamical residual effects in the inertial functions,
the constrained HFB (CHFB) plus local quasiparticle random-phase approximation (LQRPA) was proposed in Ref.~\cite{hinohara10}.
Practical applications of the CHFB + LQRPA in the $\beta$--$\gamma$ plane 
were performed only with the semi-realistic pairing-plus-quadrupole (P\,+\,Q) Hamiltonian \cite{hinohara10,sato11,hinohara11,hinohara12,sato12}.
Note that, for axially symmetric shapes without the $\gamma$ degree of freedom, 
there have been a few attempts with the Skyrme EDF \cite{yoshida11,washiyama21}.
These studies showed the importance of the dynamical residual effects in the inertial functions.


Our goal is to construct the 5DCH for the Skyrme EDF
with the inertial functions including the dynamical residual effects.
To overcome the numerical difficulties,
%
we employ the finite-amplitude method (FAM) \cite{nakatsukasa07}
that gives the response to an external one-body field.
The result of the FAM is equivalent to that of the QRPA linear-response calculation,
while the computational cost of the FAM is significantly lower than that of the QRPA.
%
The FAM has been applied to various objectives \cite{inakura09,avogadro11,stoitsov11,liang13,niksic13,hinohara13,mustonen14,pei14,kortelainen15,hinohara15b,hinohara16,wen16,washiyama17,sun17,hinohara22,sasaki22}.
The formulation with the FAM for the inertia associated with zero-energy modes
was given in Ref.~\cite{hinohara15b},
and applied to the inertia for pairing rotations \cite{hinohara15b,hinohara16} and that for spatial rotations \cite{kortelainen18,washiyama18}. 
The present study is an extension of the methods developed in Refs.~\cite{hinohara15b,washiyama21}
to the inertial functions in the 5DCH
with $\beta$ and $\gamma$.

\textit{5DCH method.}
The five-dimensional quadrupole collective Hamiltonian is given as \cite{bohr}
\begin{align}\label{eq:bohr_hamiltonian}
    H_{\text{coll}} &= T_{\text{vib}} + T_{\text{rot}} +V(\beta,\gamma),\\
        T_{\text{vib}} &= \frac{1}{2}D_{\beta\beta}(\beta,\gamma) \dot{\beta}^2
    + D_{\beta\gamma}(\beta,\gamma) \dot{\beta}\dot{\gamma} + \frac{1}{2} D_{\gamma\gamma}(\beta,\gamma)\dot{\gamma}^2, \label{eq:kinetic_vib}\\
    T_{\text{rot}} & = \frac{1}{2}\sum_{k=1}^3 \mathcal{J}_k(\beta,\gamma)(\omega^{\text{rot}}_k)^2, \label{eq:kinetic_rot}
\end{align}
where the collective potential $V$ and all the inertial functions appearing in
the vibrational $T_{\rm vib}$ and rotational $T_{\rm rot}$ kinetic energies
depend on $\beta$ and $\gamma$.
The inertial functions $D_{\beta\beta}$, $D_{\beta\gamma}$, $D_{\gamma\gamma}$, and 
$\mathcal{J}_k(\beta,\gamma) = 4\beta^2 D_k(\beta,\gamma) \sin^2(\gamma-2\pi k/3)$
denote the vibrational masses and the rotational moments of inertia, respectively. $\omega^{\text{rot}}_k$ are the rotational angular velocities in the body-fixed frame.
We use the Pauli prescription to quantize the Hamiltonian \eqref{eq:bohr_hamiltonian} and obtain the excitation energies and collective wave functions.
More details of the 5DCH method can be found in Refs.~\cite{prochniak09,hinohara10,Matsuyanagi:2016gyp}.

The collective potential is given by the energy
at the state $\ket{\phi(\beta,\gamma)}$
obtained by solving the CHFB equation
with constraints on 
the mass quadrupole operators 
$\hat{Q}_{20} = \sum_{i=1}^A r_i^2Y_{20}(\hat{r}_i)$ and $\hat{Q}_{22} = \sum_{i=1}^A r_i^2[Y_{22}(\hat{r}_i) + Y_{2-2}(\hat{r}_i)]/\sqrt{2}$.
The quadrupole deformation parameters are written as 
$\beta\cos\gamma = \eta Q_{20} = \eta \braket{\phi(\beta,\gamma)|\hat{Q}_{20}|\phi(\beta,\gamma)}$ and
$\beta\sin\gamma = \eta Q_{22} = \eta \braket{\phi(\beta,\gamma)|\hat{Q}_{22}|\phi(\beta,\gamma)}$
with $\eta = 4\pi/(3 R^2 A)$ and $R=1.2 A^{1/3}~\text{fm}$ of the mass number $A$.

\textit{Inertial functions.}
The CHFB+LQRPA inertial functions in the $\beta$--$\gamma$ plane
are given in Ref. \cite{hinohara10}. 
Although the LQRPA are defined at each CHFB state $\ket{\phi(\beta,\gamma)}$,
for simplicity, 
we abbreviate $\ket{\phi(\beta,\gamma)}$ as $\ket{\phi}$
in the following.

We employ the FAM \cite{nakatsukasa07} for a solution of the LQRPA equation for the Skyrme EDF,
which provides the $X$ and $Y$ amplitudes induced by an external field $\hat{F}$ 
at a given frequency $\omega$.
Following Refs. \cite{hinohara15b,washiyama21},
the relation between the amplitudes $(X,Y)$
and the local generators $(\hat{Q}_i,\hat{P}_i)$ 
for the $i$th normal mode with the eigenfrequency $\Omega_i$
is given in their two-quasiparticle (2qp) components as
\begin{subequations}\label{eq:FAMamplitude-xy-reduced} 
\begin{align}
  X_{\mu\nu}(\omega;\hat{F}) =& \sum_i \frac{1}{\omega^2 -\Omega^2_i}
   (P_{\mu\nu}^i + i\omega Q_{\mu\nu}^i)
        p_i(\hat{F}) ,\\
  Y_{\mu\nu}(\omega;\hat{F}) =& \sum_i \frac{1}{\omega^2 -\Omega^2_i}
  (-P_{\mu\nu}^{i*} - i\omega Q_{\mu\nu}^{i*})
        p_i(\hat{F}) ,
\end{align} \end{subequations}
where we fix the normalization of $(\hat{Q}_i,\hat{P}_i)$ 
to make the local inertial mass $M_i=1$ \cite{washiyama21}. 
Here, $\hat{F}$, $\hat{Q}_i$, and $\hat{P}_i$ are all Hermitian.
Their 2qp components $F_{\mu\nu}$ and $Q_{\mu\nu}^i$ are chosen to be real,
while $P_{\mu\nu}^i$ are pure imaginary.
In Eq.~\eqref{eq:FAMamplitude-xy-reduced}, 
the pure imaginary quantities 
$p_i(\hat{F})$ are given by
\begin{equation}
  p_i(\hat{F}) \equiv \bra{\phi} [\hat{P}_i, \hat{F}]\ket{\phi}
  = \sum_{\mu<\nu} \left(P_{\mu\nu}^{i*} -P_{\mu\nu}^{i} \right)F_{\mu\nu}. \label{eq:transitionP}
\end{equation}
The FAM response function,
$S_{\hat{F}^\prime,\hat{F}}(\omega)$,
for Hermitian and real operators, $\hat{F}$ and $\hat{F^\prime}$,
is defined as
\begin{align}
  S_{\hat{F}^\prime,\hat{F}}(\omega) &\equiv \sum_{\mu < \nu} \left[F_{\mu\nu}^{\prime 20*} X_{\mu\nu}(\omega;\hat{F})+ F_{\mu\nu}^{\prime 02*} Y_{\mu\nu}(\omega;\hat{F})\right]  \label{eq:FAMresponse} \\
  &=\sum_i \frac{1}{\omega^2 -\Omega^2_i}
 p_i(\hat{F}) p^*_i(\hat{F}^\prime).
  \label{eq:QRPAresponse}
\end{align}
Then, a contour integration of Eq.~\eqref{eq:QRPAresponse} with a contour $C_i$ that encloses the pole $\omega=\Omega_i$ 
in the complex energy plane,
\begin{equation}\label{eq:contour_integration}
  \frac{1}{2\pi i} \oint_{C_i} \omega  S_{\hat{F}^\prime,\hat{F}}(\omega) d\omega
  = \frac{1}{2}  p_i(\hat{F}) p^*_i(\hat{F}^\prime)
\end{equation}
gives $p_i(\hat{Q}_{20})$ and $p_i(\hat{Q}_{22})$
by proper choices of the operators $F$ and $F'$.

We select two LQRPA normal modes for the collective coordinates $q_i$ ($i=1,2$).
The prescription for this selection is given in Ref.~\cite{hinohara10}
(see also {\it Numerical procedure} below).
The kinetic energy
of the LQRPA normal modes in the diagonal form is rewritten
in terms of the collective variables $Q_{2m}$ ($m=0,2$) as 
\begin{equation}
    T_{\text{vib}} =  \frac{1}{2}\sum_{i=1,2}\dot{q}_i^2 = 
    \frac{1}{2}\sum_{m,n=0,2}\mathcal{M}_{mn}\dot{Q}_{2m}\dot{Q}_{2n},
\end{equation}
where the vibrational inertia tensor $\mathcal{M}_{mn}$
is obtained by 
\begin{equation}\label{eq:vibrational-mass}
\mathcal{M}_{mn} =
\sum_{i=1,2} \frac{\partial q_i}{\partial Q_{2m}} \frac{\partial q_i}{\partial Q_{2n}}.
\end{equation}
The inverses of these partial derivatives are evaluated as
\begin{align}\label{eq:partial-derivative}
\frac{\partial Q_{2m}}{\partial q_i} 
&=  \frac{\partial }{\partial q_i} \langle \phi | \hat{Q}_{2m}|\phi\rangle  \notag \\
&= \langle \phi | [\hat{Q}_{2m}, \frac{1}{i}\hat{P}_{i}]|\phi\rangle
=i p_i(\hat{Q}_{2m}).
\end{align}
Thus, the inertia tensor \eqref{eq:vibrational-mass}
is obtained by the FAM calculation of
$p_i(\hat{Q}_{2m})$ in Eq. (\ref{eq:QRPAresponse})
through Eq.~\eqref{eq:partial-derivative}.

With the relation between ($\beta$, $\gamma$) and ($Q_{20}$, $Q_{22}$),
the vibrational masses $D_{\beta\beta}$, $D_{\beta\gamma}$, and $D_{\gamma\gamma}$ are obtained from $\mathcal{M}_{00}$, $\mathcal{M}_{02}$, and $\mathcal{M}_{22}$ in Eq.~\eqref{eq:vibrational-mass} \cite{hinohara10}.
Note that the formulation given above can be extended to cases with more than two collective variables.

For the rotational moments of inertia, 
the Thouless-Valatin rotational moment of inertia $\mathcal{J}_k$
at the CHFB state $\ket{\phi}$
is evaluated from the FAM strength function at zero energy as
$S_{\hat{J}_k,\hat{J}_k}(\omega=0) = -\mathcal{J}_k$,
where $\hat{J}_k$ is the angular momentum operator~\cite{hinohara15b}.

\textit{Numerical procedure.}
We solve the CHFB + LQRPA equations following Ref.~\cite{washiyama21}.
We calculate the vibrational masses with the FAM-LQRPA in two steps.
First, 
to find peaks in the strength distribution,
we solve the FAM equations with the external fields $\hat{Q}_{20}$ and $\hat{Q}_{22}$
at $0 \le |\omega| \le 4\,\text{MeV}$,
in both real and imaginary $\omega$
with a smearing width of 0.01~MeV.
The peak position should correspond to the LQRPA poles $\Omega_i$.
Second, for each pole, we perform the contour integration~\eqref{eq:contour_integration} with a circle of radius 0.02~MeV discretized to eight points.
Then, we select the two most collective LQRPA modes 
following the prescription~\cite{hinohara10}
that a pair of LQRPA solutions that give the minimum value of $W =(D_{\beta\beta}D_{\gamma\gamma}-D_{\beta\gamma}^2)/\beta^2$
is selected from many LQRPA solutions.
In practice, we select several peaks with large strengths
and calculate Eq.~\eqref{eq:contour_integration} with all the combinations from the selected peaks
to find the minimum $W$.

We solve the CHFB equations with the two-basis method \cite{gall94,terasaki95} in the three-dimensional Cartesian mesh with a $(13.2\,\text{fm})^3$ box with a mesh size of 0.8\,fm.
The reflection symmetries about $x=0$, $y=0$, and $z=0$ planes %
lead to the single-particle states as eigenstates of the parity,  $z$ signature, and $y$-time simplex~\cite{bonche85,ev8,ev8new,hellemans12}.
The single-particle basis consists of 1400 neutron and 1120 proton HF-basis states, which approximately correspond to the maximum quasiparticle energy of 60~MeV for $^{110}$Pd
and give a good convergence in the CHFB and LQRPA calculations~\cite{washiyama21}.
We employ the SkM$^*$ EDF~\cite{bartel82} and the contact volume-type pairing with a pairing window of 20~MeV above and below the Fermi level as described in Ref.~\cite{ev8new}.
The pairing strengths are adjusted to reproduce the empirical neutron and proton gaps in $^{106}$Pd.
We use an equilateral triangular mesh of $\Delta \beta \approx 0.05$ 
in $0 < \beta < 0.6$ and $0^\circ < \gamma < 60^\circ$
in the $\beta$--$\gamma$ plane, 
consisting of 93 deformation points.

The numerical calculations of the FAM are performed with hybrid parallelization
(MPI + OpenMP).
For the vibrational masses,
it takes about 480 core hours to select several peaks in the strength distributions 
and 150~core hours for a contour integration
for each deformation point.
For the three rotational moments of inertia, 
it takes about 3~core hours for each deformation point.
Computing the LQRPA inertial functions is feasible in currently available computational resources.

\begin{figure}
\includegraphics[width=\linewidth,clip]{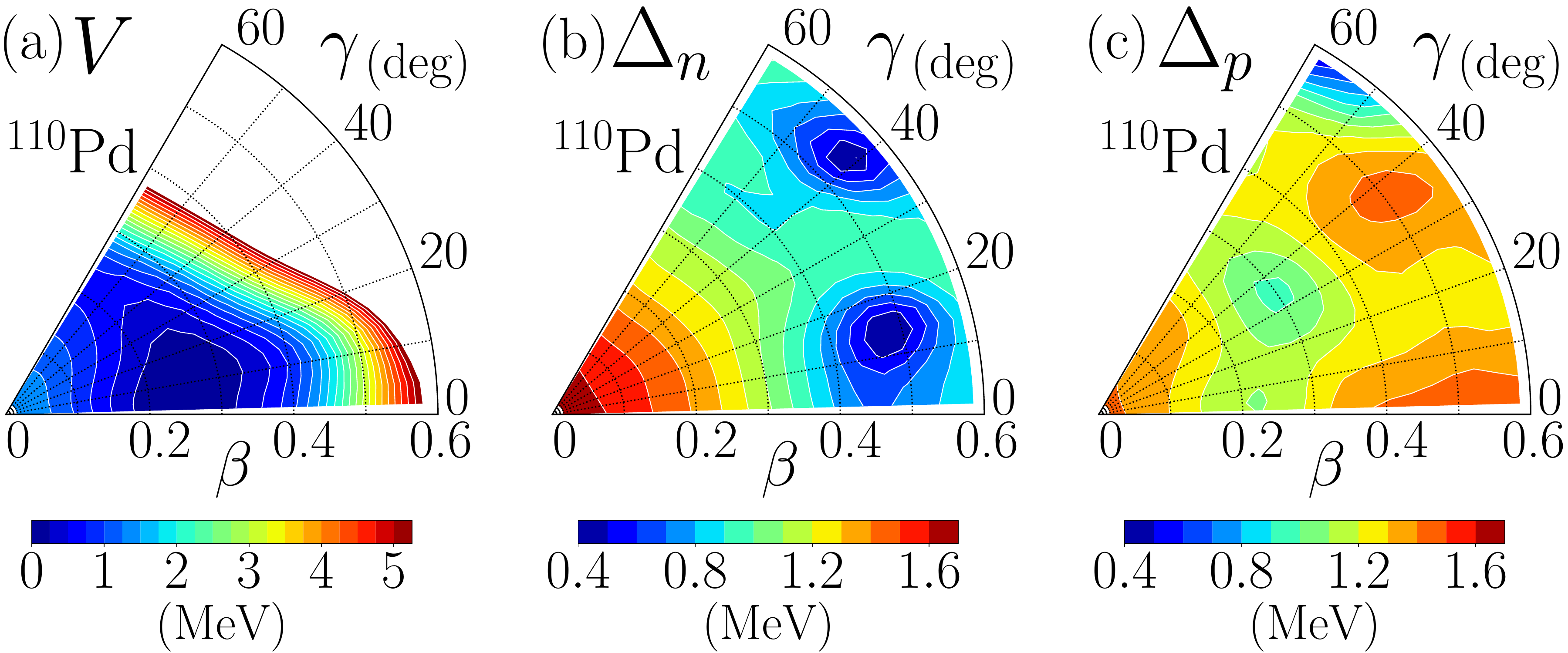}
\caption{
Potential energy surface (a) and pairing gaps for neutrons (b) and protons (c) in the $\beta$--$\gamma$ plane in $^{110}$Pd.
        }\label{fig:PES_gap_Pd110}
\end{figure}     

\textit{Results and discussions.}
Figure~\ref{fig:PES_gap_Pd110}(a) shows the calculated potential energy surface (PES) measured from the energy minimum in the $\beta$--$\gamma$ plane in $^{110}$Pd.
The shallow energy minimum is found at $\beta\approx 0.25$ and $\gamma \approx 0^\circ$.
The PES is flat in both the $\beta$ and $\gamma$ directions
with $V(\beta,\gamma)<1$~MeV in a wide region of $0.1<\beta<0.4$ 
and $0^\circ < \gamma < 60^\circ$. 
Figures~\ref{fig:PES_gap_Pd110}(b) and \ref{fig:PES_gap_Pd110}(c) show the pairing gaps for neutrons $\Delta_n$ and protons $\Delta_p$, respectively, in $^{110}$Pd. 
The pairing gap in neutrons has local minima at $\beta\approx 0.5,\gamma \approx 15^\circ$ and $\beta\approx 0.6,\gamma\approx 40^\circ$.

\begin{figure}
    \includegraphics[width=\linewidth]{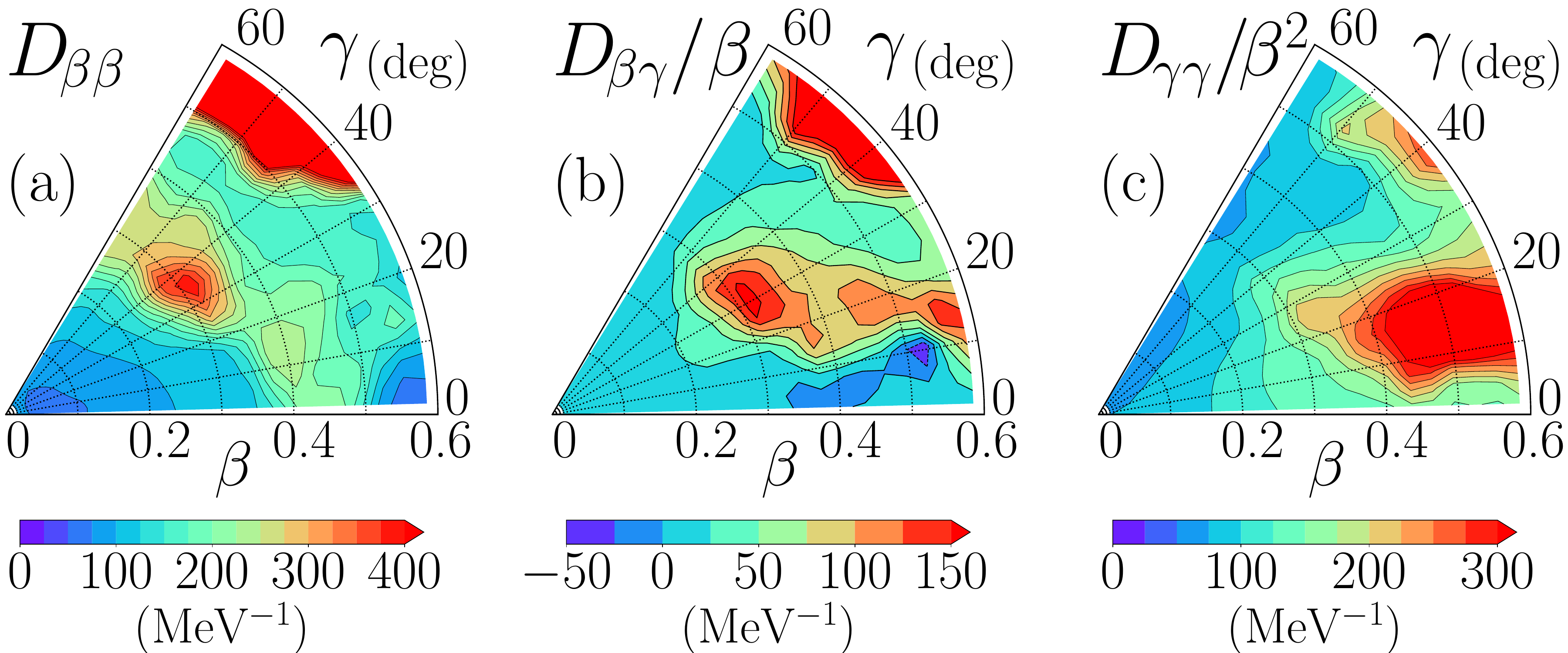} 
    \\
    \vspace{0.5em}
    \includegraphics[width=0.67\linewidth]{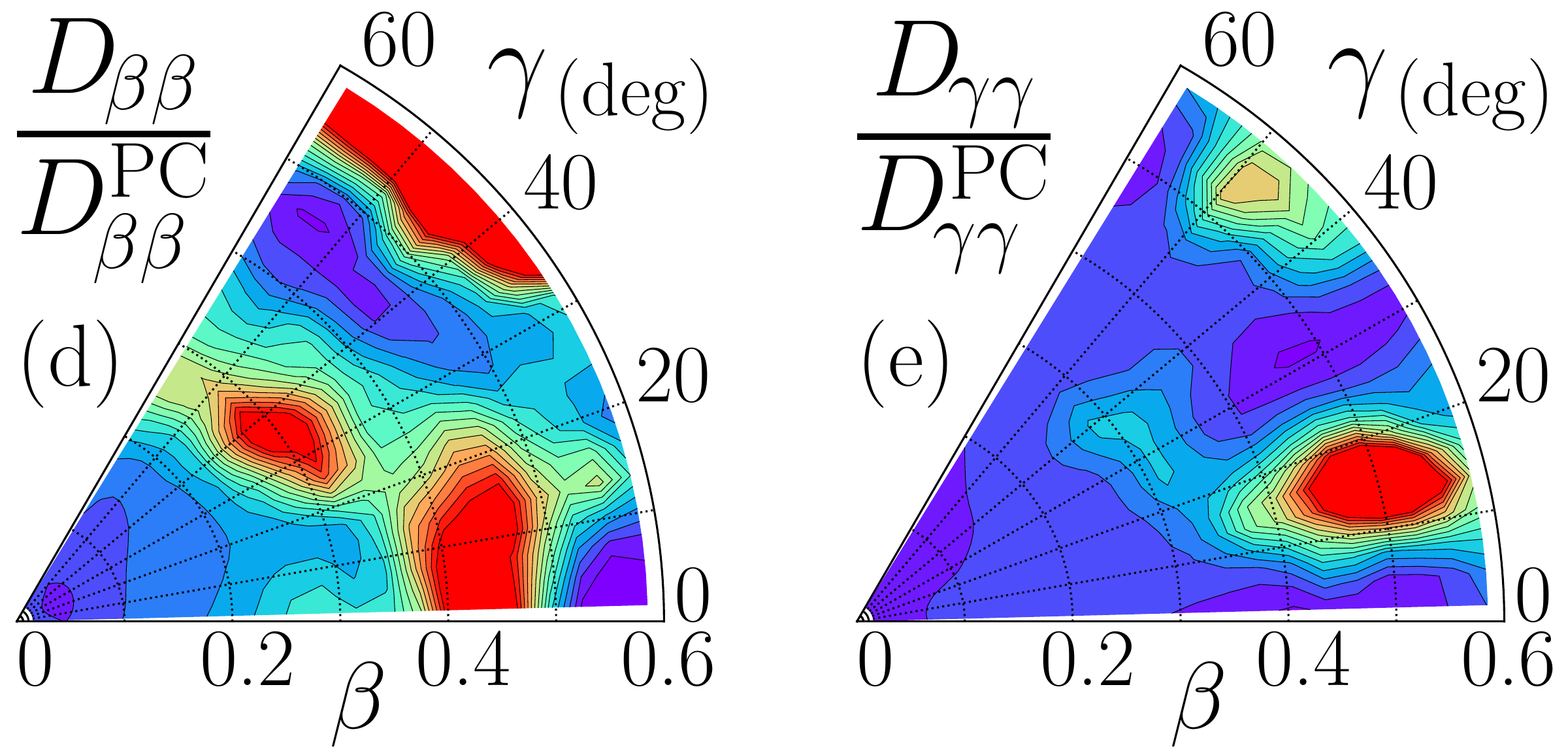} 
    \hspace{0.2em}
    \includegraphics[width=0.038\linewidth,clip]{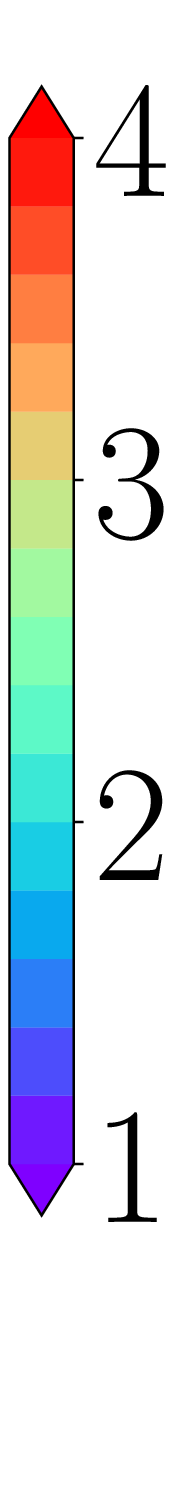}
    \caption{
    Vibrational masses of the LQRPA in ${}^{110}$Pd:
    (a) $D_{\beta\beta}$, (b) $D_{\beta\gamma}/\beta$, and (c) $D_{\gamma\gamma}/\beta^2$.
    The ratio of the LQRPA to the PC vibrational masses:
    (d) $D_{\beta\beta}/D_{\beta\beta}^{\rm PC}$ and (e)  $D_{\gamma\gamma}/D_{\gamma\gamma}^{\rm PC}$.
        }\label{fig:vibmass_Pd110}
\end{figure} 

Figure~\ref{fig:vibmass_Pd110} shows the vibrational masses $D_{\beta\beta}$ (a), $D_{\beta\gamma}/\beta$ (b), and $D_{\gamma\gamma}/\beta^2$ (c) calculated with
the CHFB + LQRPA in the $\beta$--$\gamma$ plane for ${}^{110}$Pd.
A remarkable feature is a strong variation of the vibrational masses
in the $\beta$--$\gamma$ plane. 
In particular, the vibrational masses become locally large at around $\beta = 0.5$ and $\gamma = 15^\circ$,
$\beta = 0.3$ and $\gamma = 40^\circ$, and
$\beta>0.4$ and $\gamma>40^\circ$,
at which the pairing gaps in neutrons and protons become locally small.
A correlation between the vibrational mass and pairing gap was also observed in the collective inertia in spontaneous fission~\cite{washiyama21}.

It is of significant importance to compare the vibrational masses obtained by the LQRPA with those obtained by the perturbative cranking (PC) formula \cite{girod79}, denoted as 
$D^{\text{PC}}_{\beta\beta}$,  $D^{\text{PC}}_{\beta\gamma}$, and $D^{\text{PC}}_{\gamma\gamma}$.  
Those of the PC formula have been extensively employed in the EDF-based
5DCH studies \cite{libert99,niksic09,delaroche10}. 
Figures~\ref{fig:vibmass_Pd110}(d) and (e) show the ratio of the LQRPA vibrational mass to the PC one.
At a region near the minimum of the PES ($\beta\approx 0.25$ and $\gamma \approx 0^\circ$), the ratio is 1.0--2.0. 
However, the ratio exceeds 3.0 at the regions at which $D_{\beta\beta}$ and $D_{\gamma\gamma}$ take large values.
Furthermore, the ratio shows a strong $\beta$--$\gamma$ dependence
and different properties in $D_{\beta\beta}$ and $D_{\gamma\gamma}$.
Former EDF-based 5DCH studies 
have often employed the cranking inertial functions multiplied
by a constant enhancement factor
to include the dynamical residual effects.
However, our findings clearly show that the use of the constant enhancement factor cannot be justified in the description of the vibrational masses. 
Similar enhancement 
is observed in the former LQRPA studies with the P\,+\,Q model
\cite{hinohara10,sato11,hinohara12}
and in those with the axial symmetric restriction 
\cite{yoshida11,washiyama21}.

\begin{figure}
    \includegraphics[width=\linewidth]{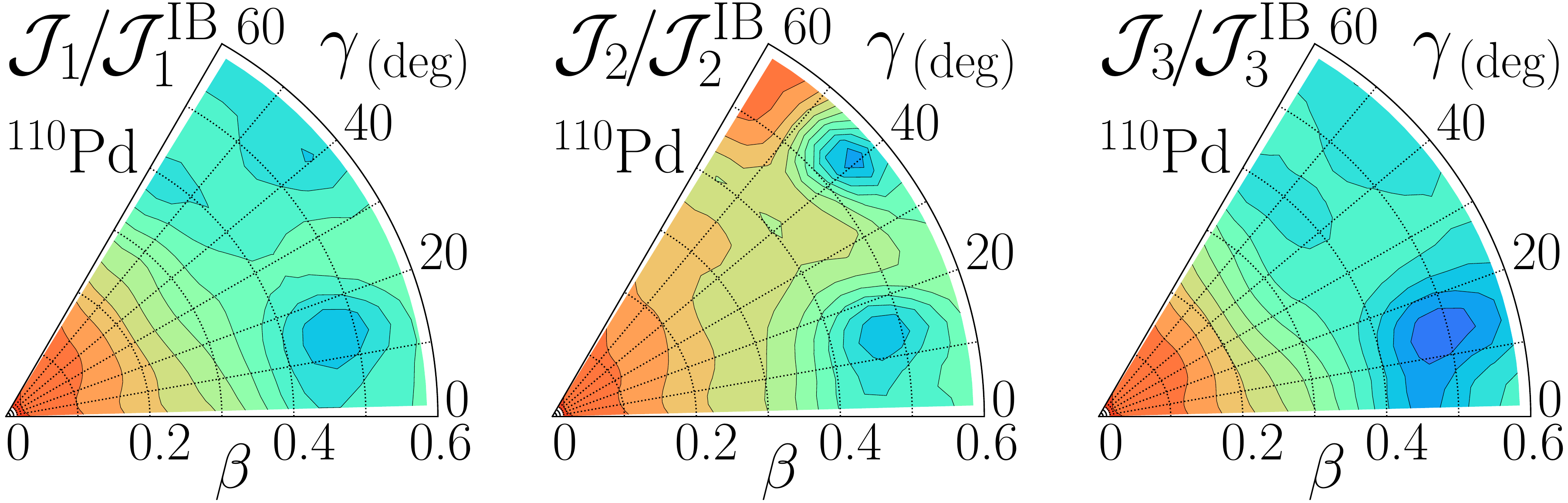} \\
    \vspace{0.2em}
     \includegraphics[width=0.33\linewidth]{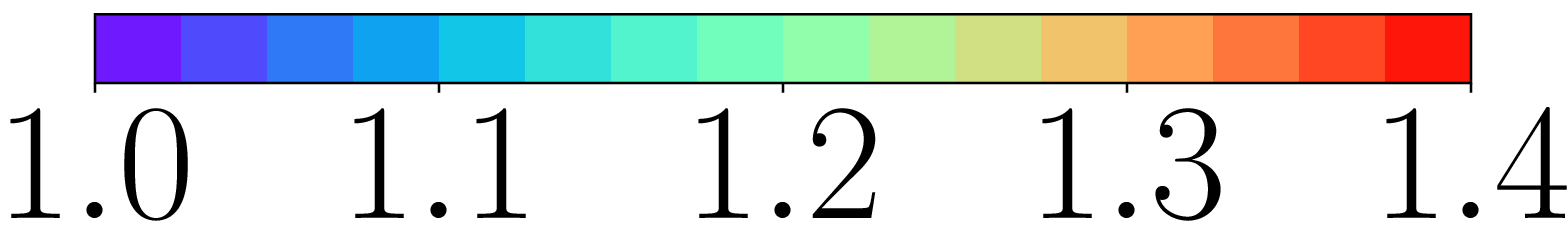}  
    \caption{
    Ratio of the LQRPA moments of inertia to the Inglis-Beliaev (IB) ones ($\mathcal{J}_k/\mathcal{J}_k^{\text{IB}}$) in the $\beta$--$\gamma$ plane in ${}^{110}$Pd.
        }\label{fig:moi_Pd110}
\end{figure}    
Figure~\ref{fig:moi_Pd110} shows
the ratios of the LQRPA to the IB cranking moments of inertia $\mathcal{J}_k^{\text{IB}}$ \cite{inglis56,beliaev61}
in the $\beta$--$\gamma$ plane for ${}^{110}$Pd.
The ratios are in a range of 1.0--1.4 and increase as $\beta$ decreases; the dynamical residual effects become larger toward the spherical shape.
The ratios become small where the pairing gap is small, 
which is opposite to the case of the vibrational masses.
The enhancement is less pronounced than that in the vibrational masses; 
this indicates the larger dynamical residual effects in the vibrational masses than in the moments of inertia.
This is consistent with the LQRPA studies with the P\,+\,Q model \cite{hinohara10,sato11,hinohara12}.

\begin{figure*}
    \includegraphics[width=0.85\linewidth]{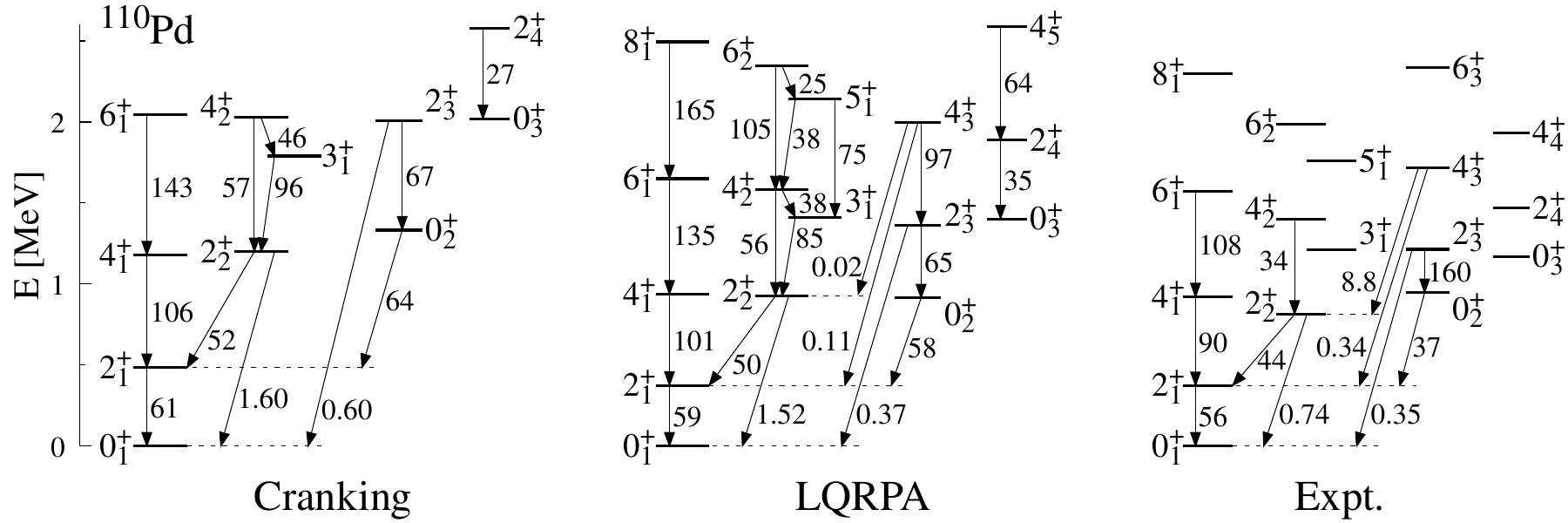}
    \caption{
    Low-lying excitation spectra and $B(E2)$ values in Weisskopf units in ${}^{110}$Pd obtained with 
    the cranking inertial functions (left), with the LQRPA ones (middle), compared with the experimental data (right) \cite{NNDC}.
    See text for details.
            }\label{fig:level_Pd110}
\end{figure*}    

Figure~\ref{fig:level_Pd110} shows excitation spectra of the 5DCH
for positive-parity $I_\alpha^+$ states 
with $I$ being the total angular momentum 
and $\alpha$ distinguishing the states with the same $I$
and $B(E2)$ values in Weisskopf units in $^{110}$Pd.
Those of the cranking inertial functions indicate the results
obtained with the PC vibrational masses and IB moments of inertia.
In the spectra, we show 
the states with $I\le 8$
and the excitation energies $E<2.6$~MeV and the $B(E2)$ values of all the intraband transitions and of the interband ones whose experimental data are available.
It is clearly seen that the excitation energies calculated with the LQRPA inertial functions are lower than those calculated with the cranking ones.
The enhancement of the LQRPA inertial functions 
lowers the ground-state rotational energies as well as the excited band-head energies.
The low-lying spectra of the LQRPA are in good agreement with the experimental data, showing a clear advantage over those with the cranking inertial functions.
However, some discrepancies from the experimental spectra remain particularly at excited bands.
For instance,
we overestimate the level spacing between $0^+$ and $2^+$ for the $0_2^+$ and the $0_3^+$ bands. 

The $B(E2)$ values in the ground-state rotational band and 
$B(E2; 0_2^+ \to 2_1^+)$ and $B(E2; 2_2^+ \to 2_1^+)$ 
agree well with the experimental ones.
The $B(E2;2_3^+ \to 0_2^+)$ value with the LQRPA underestimates the experimental data,
which is related to the overestimation of the level spacing of the $0_2^+$ and the $2_3^+$ bands.
Overall, the dynamical residual effects lead to a better agreement in the property of the low-lying spectra.
In contrast to the excitation energies,
the $B(E2)$ values calculated with the LQRPA and the cranking inertial functions are similar to each other.
It may suggest that the contribution of the dynamical residual effects is
more important for the energy properties than for the wave functions,
although it is dangerous to generalize the results for other nuclei.

\begin{figure}
     \includegraphics[width=0.935\linewidth]{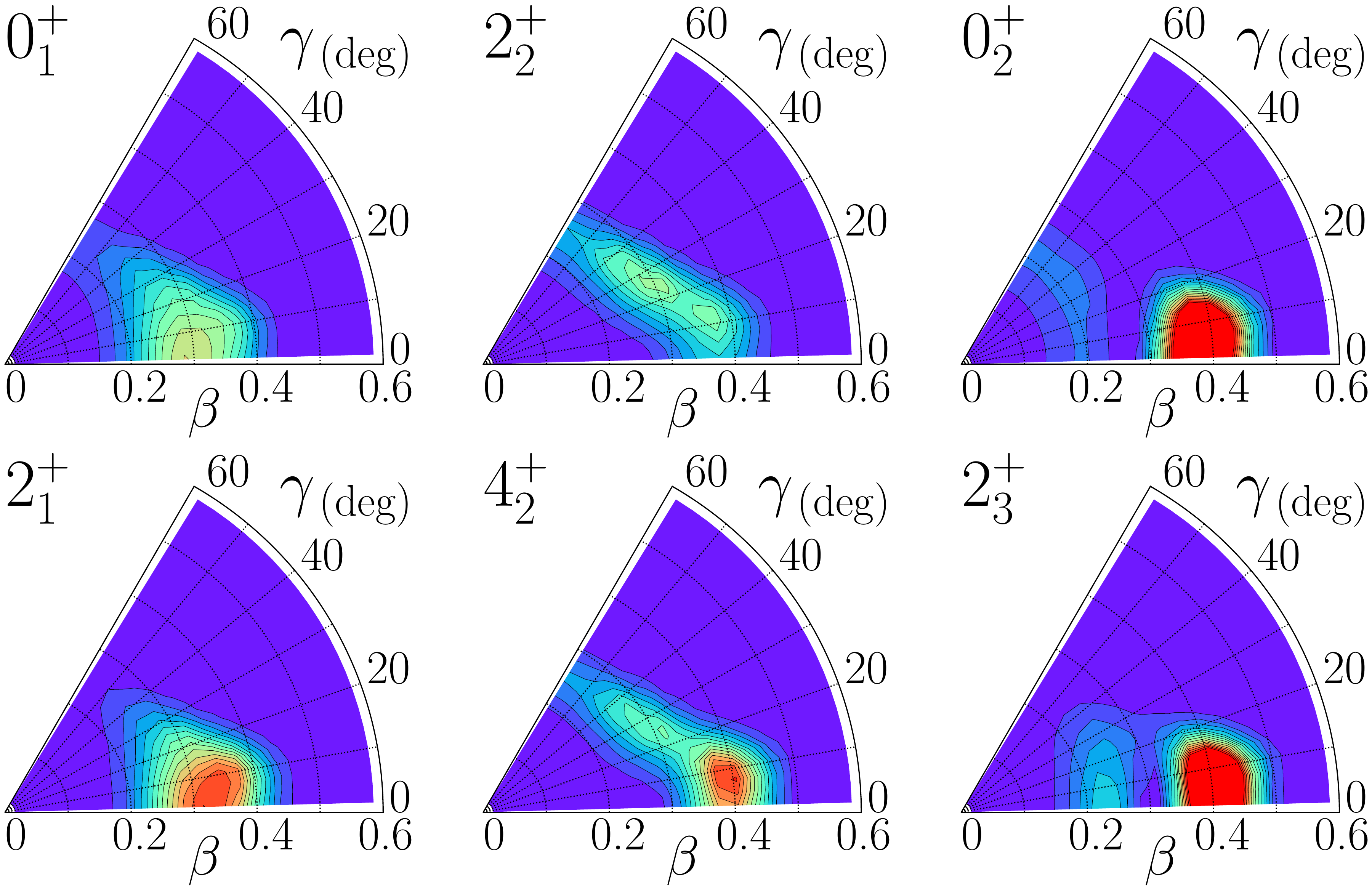}
     \includegraphics[width=0.05\linewidth]{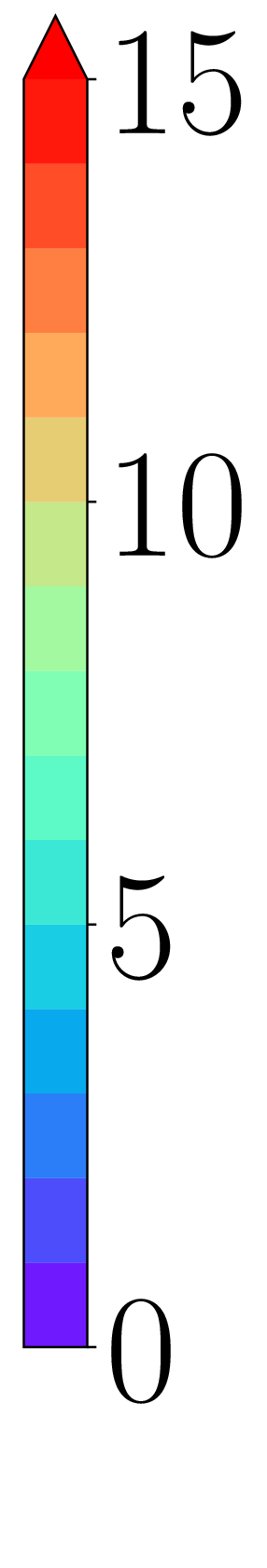}
    \caption{ 
    Vibrational wavefunctions $|\Phi_{\alpha I}(\beta,\gamma)|^2$ of $I_\alpha^+$ states
    multiplied by $\beta^4\sqrt{W(\beta,\gamma)R(\beta,\gamma)}$ in $^{110}$Pd. }
    \label{fig:vibrational_wavefunction}
\end{figure}

Finally, we discuss the property of shape fluctuations in the low-lying spectra.
Figure~\ref{fig:vibrational_wavefunction} shows the vibrational wave functions of $I_{\alpha}^{+}$ states
as $|\Phi_{\alpha I}(\beta,\gamma)|^2 \equiv \sum_{K}|\Phi_{\alpha I K}(\beta,\gamma)|^2$
with $K$ being the $z$ component of $I$ in the body-fixed frame.
We multiply the wave functions by  $\beta^4\sqrt{W(\beta,\gamma)R(\beta,\gamma)}$ 
with $W=(D_{\beta\beta}D_{\gamma\gamma}-D_{\beta\gamma}^2)/\beta^2$ and $R=D_1 D_2 D_3$ 
from the volume element in the normalization of the vibrational wave functions.
The $0_1^+$ wave function has a broad peak around $\beta\approx 0.3$, $\gamma \approx 0^\circ$, which is close to the minimum of the PES, and spreads over along both the $\beta$ and $\gamma$ directions.
This shows that the ground state has a large-amplitude shape fluctuation in the $\beta$--$\gamma$ plane.
The $2_1^+$ state shows a similar structure but is more localized around $\beta\approx 0.3$, $\gamma\approx 0^\circ$ than the $0_1^+$ state is.
The $2_2^+$ wave function has a broad peak at a triaxial shape $\beta\approx 0.3$ and $\gamma\approx 30^\circ$ and spreads over along the $\gamma$ direction.
The $4_2^+$ wave function has a feature similar to the $2_2^+$ one, which indicates that the $4_2^+$ state is a member of the $2_2^+$ band and is localized more in the prolate side.
The $0_2^+$ and $2_3^+$ wave functions show a 
feature of the $\beta$ vibration, which has a node along the $\beta$ direction.
However, the component at around $\beta\approx 0.2$ spreads along the $\gamma$ direction.
This reflects a $\gamma$-soft character. 
These collective wave functions clearly show the importance of including the triaxial degree of freedom in the 5DCH
for a transitional nucleus $^{110}$Pd.

\textit{Summary.}
We have developed a method of calculating 
the inertial functions of the 5DCH, 
the vibrational masses and the rotational moments of inertia,
by the LQRPA with the Skyrme EDF in the $\beta$--$\gamma$ plane.
The method 
can take into account the time-odd mean fields
as the dynamical residual effects in the inertial functions.
We constructed the 5DCH with the LQRPA inertial functions
and described low-lying states in a transitional nucleus $^{110}$Pd.
The dynamical residual effects increase both the vibrational masses 
and rotational moments of inertia, 
compared with those within the cranking formula.
This enhancement strongly depends on both $\beta$ and $\gamma$, 
which indicates an insufficient treatment of the constant enhancement factor to the cranking inertial functions employed in former EDF-based 5DCH studies.
A good agreement with the experimental data is achieved for low-lying spectra.
The vibrational wave functions in the low-lying states
show significant shape fluctuations 
in the $\beta$--$\gamma$ plane.

The present study shows the feasibility
of performing computations of the LQRPA for the inertial functions with the present computational resources.
However, 
systematic calculations of the 5DCH method across the nuclear chart will need huge computational costs.
Computational costs are expected to be further reduced by using a recent development of the reduced basis method for the FAM~\cite{hinohara23}.

\textit{Acknowledgments.}
The authors thank K. Yaoita for providing his numerical code for solving the collective Schr\"odinger equation.
This work was supported in part by the JSPS KAKENHI (Grant No.~JP19KK0343, No.~JP20K03964, No.~JP22H04569, and No.~JP23H01167).
Numerical calculations were performed using computational resources of Wisteria/BDEC-01 Odyssey (the University of Tokyo), provided by the Multidisciplinary Cooperative Research Program in the Center for Computational Sciences, University of Tsukuba.

%

\end{document}